\documentstyle[aps,multicol,prb,epsf]{revtex}

\begin{document}
\title{Spin-orbit coupling effect on quantum Hall ferromagnets with vanishing
Zeeman energy}
\author{Vladimir I. Fal'ko$^{\ast }$ and S.V. Iordanskii$^{\ddagger \ast }$}
\address{$^{\ast }$ School of Physics and Chemistry, Lancaster University, LA1 4YB, UK%
\\
$^{\ddagger }$ Landau Institute for Theoretical Physics, Kosygina 1, Moscow,
Russia}
\date{today}
\maketitle

\begin{abstract}
We present the phase diagram of a ferromagnetic $\nu =2N+1$ quantum Hall
effect liquid in a narrow quantum well with vanishing single-particle Zeeman
splitting, $\varepsilon _{{\rm Z}}$ and pronounced spin-orbit coupling. Upon
decreasing $\varepsilon _{{\rm Z}}$, the spin-polarization field of a liquid
takes, first, the easy-axis configuration, followed by the formation of a
helical state, which affects the transport and NMR properties of a liquid
and the form of topological defects in it.
\end{abstract}


\ifpreprintsty\tightenlines \else 
\begin{multicols}{2} \fi

The concept of quantum Hall ferromagnets (QHF) \cite{Girvin,Lee,MacDonald}
is now broadly accepted as an approach to treat the 2D electron systems in
the vicinity of odd-integer filling factors, $\nu =2N+1$, in particular, due
to the success in the theory \cite{MacDonald,Sondhi} and experimental
observations \cite{Eisenstein,Barrett,Aifer} of quantum Hall effect
skyrmions, which are electrically charged topological textures in the
ferromagnetic order parameter in two dimensions. The attempts to stabilize
skyrmions \cite{MacDonald,Sondhi}, as compared to electrons and holes on the
top of a filled spin-split Landau level (LL) have triggered the studies \cite
{Maude,Leadley} of semiconductor structures with a reduced value of the
single-electron Zeeman energy, $\varepsilon _{{\rm Z}}$.

The latter is possible to realize using GaAs/AlGaAs structures, due to a
strong spin-orbit (SO) coupling in this zinc-blend-type semiconductor \cite
{SOTh,Dresselhaus}. In particular, the conduction band electron Lande
factor, $g$ in GaAs may reduce its absolute value or even change sign\ under
a hydrostatic pressure, and also due to the electron confinement in a narrow
quantum well (QW). Both theory \cite{SOTh2} and experiment \cite
{Harley,Ruhle} agree on that in a non-strained $55$\AA -wide GaAs/Al$_{.33}$%
Ga$_{.67}$As QW, $g=0$, and the value $g=-0.1$ has been\ recently measured
in a non-strained $w=68$\AA\ QW \cite{Maude}. As a result, a narrow quantum
well is a system, where Zeeman energy can be swept through zero, and where
the tendency of interacting electrons at $\nu =\left( 2N+1\right) $ to form
a ferromagnetic state (with a large exchange energy $\Im \gg \varepsilon _{%
{\rm Z}}$) polarized along the external magnetic field confronts the effects
of the SO coupling \cite{BychRashba,Rossler,Jusserand} itself, as an
alternative source for choosing the spin-polarization direction for 2D
electrons.

In GaAs/AlGaAs quantum wells grown along the $\left[ 001\right] $-facet, the
effective two-dimensional SO coupling has the form of $H_{{\rm so}}=%
\widetilde{u}_{{\rm so}}\left( p_{x}\sigma ^{x}-p_{y}\sigma ^{y}\right) +u_{%
{\rm so}}\epsilon ^{ijz}p_{i}\sigma ^{j}$, where ${\bf p}=-i\nabla -e{\bf A/}%
c$ is the electron momentum (${\bf A=(}0,xB_{z},0{\bf )}$), $\sigma _{j}$
are Pauli matrices, and $\hbar =1$. The first term in $H_{{\rm so}}$
reflects the lack of inversion symmetry in the well grown along the $\left[
001\right] $ facet with square 2D lattice symmetry and comes from the $%
\left( \gamma p^{3}\sigma \right) $ non-parabolicity in the conduction band
of bulk GaAs \cite{Dresselhaus,Rossler}, so that $\widetilde{u}_{{\rm so}%
}=\gamma \left\langle p_{z}^{2}\right\rangle $. Experimentally \cite
{Jusserand} and theoretically \cite{Rossler}\ found values of $\gamma $
range from $\gamma =27$eV\AA $^{3}$ to $22$eV\AA $^{3}$. The second SO term
is due to the quantum well potential asymmetry \cite{BychRashba}. In a
narrow QW, with $\psi =2^{-1/2}\sin (z\pi /w)$ transverse part of the
electron wave function, $\widetilde{u}_{{\rm so}}=\gamma \left( \pi
/w\right) ^{2}\gg u_{{\rm so}}$ \cite{Jusserand}. (Directions for $x$- and $y
$-axes\ are chosen in such a way that $\widetilde{u}_{{\rm so}},u_{{\rm so}%
}>0$.)

In the present paper, we study the effect of spin-orbit coupling on the $\nu
=\left( 2N+1\right) $ QHF formed by electrons with a vanishing
single-particle Zeeman energy in a narrow QW in a perpendicular magnetic
field. In this analysis, we address the properties of narrow wells, since
they have a more prominent SO coupling in the 2D electron Hamiltonian. In a
high magnetic field providing $\omega _{{\rm c}}=\left| eB_{z}\right| /mc\gg 
\widetilde{u}_{{\rm so}}/\lambda $ ($\lambda =\sqrt{c/\left| eB_{z}\right| }$%
), a weak SO coupling does not alter Landau quantization. However, it
affects the evolution of ferromagnetic properties of a quantum Hall effect
liquid upon sweeping $\varepsilon _{{\rm Z}}$ through zero. As in ordinary
ferromagnets, SO coupling results in the crystalline anisotropy field \cite
{Advice}, that deflects spin polarization ${\bf n}$ from alignement in the
magnetic field direction, ${\bf l}_{z}$. As a result, a liquid with $\left|
\varepsilon _{{\rm Z}}\right| <\varepsilon _{{\rm Z}}^{{\rm e}}=2\left( 
\widetilde{u}_{{\rm so}}/\lambda \omega _{{\rm c}}\right) ^{2}\Im _{N+1}$\
takes one of two two-fold degenerate easy-axis magnetic states, which
generates a source for formation of a domain structure in a QHF. Moreover,
as a feature of the SO coupling in the 2D electron system lacking inversion
symmetry, for a tinier splitting, $\left| \varepsilon _{{\rm Z}}\right|
<\varepsilon _{{\rm Z}}^{{\rm h}}=\sqrt{2}\left( \widetilde{u}_{{\rm so}%
}/\lambda \omega _{{\rm c}}\right) ^{2}\Im _{N+1}$, the spin-polarization
acquires a helically twisted texture with a mesoscopic-scale period, ${\cal L%
}/\lambda \approx 3.25\omega _{{\rm c}}\lambda /\widetilde{u}_{{\rm so}}$,
in the $\left[ 1\bar{1}0\right] $ crystallographic direction. The transition
of a liquid into the helical state may manifest itself in a change of the
NMR-lineshape from the QW structure, or in the anisotropy of dissipative
transport characteristics of a QHF. We also discuss topological defects in
the helical state, $\pm e/2$-charged dislocations, which, in pairs,
constitute skyrmions.

The recipe \cite{MacDonald,Sondhi} of describing smoothly textured QHF's at
odd-integer filling factors is to use the 2D sigma-model, which operates
with the energy functional, $\Phi \left\{ {\bf n}({\bf x})\right\} $ of 2D
electrons expressed in terms of their local excess spin polarization field $%
{\bf n(r},t{\bf )}$ (${\bf |n|=}1$). Locally, the QHF can be viewed as a
liquid of electrons which fully occupy $\left( N+1\right) $ LL's with spin
parallel to ${\bf n}$, and $N$ LL's with antiparallel spins. This assumes
the existence of a local unitary spin-transformation, $U({\bf r})$ of
electron wave functions, which reduces the filling of the $(N+1)$-st LL to a
complete occupation of only states 'up' with respect to the locally
determined axis ${\bf n(r)}$, and which is related to the spin-polarization
field as $n^{i}={\rm Tr}(s^{i}\Lambda )$, where $s^{i}({\bf r})=U({\bf r}%
)\sigma ^{i}U^{\dagger }({\bf r})$, and $\Lambda =\left( 1+\sigma
^{z}\right) /2$ is the electron spin-density matrix in the rotated frame.
Such a liquid retains incompressibility, which is guaranteed by a large
exchange energy gap, $\Im $. The derivation of a sigma-model consists in the
use of Hubbard-Stratonovich transformation and the saddle-point
self-consistency equation, as a method to obtain an expansion of the
thermodynamic potential of a liquid over small gradients of a polarization
field, or, equivalently, over the matrix $\overrightarrow{\Omega }({\bf r}%
)=U({\bf r})\left( -i\nabla U^{\dagger }({\bf r})\right) \equiv \left(
i\nabla U({\bf r})\right) U^{\dagger }({\bf r})$. The latter matrix appears
in a local perturbation to the single-particle Hamiltonian written in the
rotated spin-frame, 
\[
H_{\Omega }=i\omega _{{\rm c}}\left( a_{+}\Omega _{-}-\Omega
_{+}a_{-}\right) +\left( \omega _{{\rm c}}/2\right) \left[ \vec{\Omega}%
^{2}-\alpha \nabla \times \vec{\Omega}\right] ,
\]
where $a_{\mp }=\pm (ip_{x}\mp \alpha p_{y})/\sqrt{2}$ are the inter-LL
operators, $\alpha =eB_{z}/\left| eB_{z}\right| $ indicates the direction of
the cyclotron rotation of carriers, and $\Omega _{\pm }=(\Omega _{x}\mp
i\alpha \Omega _{y})/\sqrt{2}$. Further procedure (equivalent to the
Hartree-Fock approximation) consists in\ a perturbative expansion of the
saddle-point Hubbard-Stratonovich action (obtained for a given
single-particle Hamiltonian, which includes $H_{\Omega }+H_{{\rm so}}$) up
to the second order both in $\overrightarrow{\Omega }$ and the SO coupling, $%
H_{{\rm so}}$ (assuming that $\widetilde{u}_{{\rm so}}/\lambda \ll \Im $).
In the local spin-frame, 
\begin{eqnarray*}
H_{{\rm so}} &=&\left( \widetilde{u}_{{\rm so}}/\lambda \right) \left[
i(a_{+}s^{+}-s^{-}a_{-})+\Omega _{+}s^{+}+s^{-}\Omega _{-}\right]  \\
&&+\left( \alpha u_{{\rm so}}/\lambda \right) \left[
a_{+}s^{-}+s^{+}a_{-}-i(\Omega _{+}s^{-}-s^{+}\Omega _{-})\right] ,
\end{eqnarray*}
where $s^{\pm }=(s^{x}\mp \alpha is^{y})/\sqrt{2}$, and the relevant part of 
$H_{{\rm so}}$ is off-diagonal with respect to the LL-number \cite
{FalkoPRL93}.

The calculation, which leads us to the sheet density of thermodynamic
potential of a QHF, $\Phi \left\{ {\bf n}({\bf x})\right\} $ differs from
earlier Hartree-Fock calculations \cite{MacDonald,FIP} only by taking into
account the SO-coupling term $H_{{\rm so}}$, along with $H_{\Omega }$. It
results in

\ifpreprintsty
\else 
\end{multicols}\vspace*{-3.5ex}{\tiny \noindent 
\begin{tabular}[t]{c|}
\parbox{0.493\hsize}{~} \\ \hline
\end{tabular}
} \fi

\begin{equation}
\Phi =\frac{\Im _{N+1}}{2\pi }\left\{ \sum_{\beta }\frac{\left( \nabla
n^{\beta }\right) ^{2}}{8}-\left[ \frac{\widetilde{u}_{{\rm so}}^{2}+u_{{\rm %
so}}^{2}}{2\omega _{{\rm c}}^{2}\lambda ^{4}}{\bf n}_{\Vert }^{2}+\frac{2%
\widetilde{u}_{{\rm so}}u_{{\rm so}}}{\omega _{{\rm c}}^{2}\lambda ^{4}}%
n^{x}n^{y}\right] +\frac{\varepsilon _{{\rm Z}}n^{z}}{2\lambda ^{2}}-\frac{%
\widetilde{u}_{{\rm so}}\widetilde{{\bf n}}_{\Vert }-u_{{\rm so}}{\bf n}%
_{\Vert }}{\omega _{{\rm c}}\lambda ^{2}}\cdot \nabla n^{z}\right\} +E_{{\rm %
sk}},  \label{Phi}
\end{equation}
where ${\bf n}_{\Vert }$ is the planar component of the spin-polarization
field, and $\widetilde{n}^{x}=n^{y}$, $\widetilde{n}^{y}=n^{x}$. To obtain $%
\Phi $ in Eq. (\ref{Phi}), we have extended perturbative expansion up to the
terms $\Im \nabla ^{2},$ $\Im \left( \widetilde{u}_{{\rm so}}/\lambda \omega
_{{\rm c}}\right) \nabla $ and $\Im \left( \widetilde{u}_{{\rm so}}/\lambda
\omega _{{\rm c}}\right) ^{2}$. The exchange-factor, 
\[
\Im _{N+1}=\int_{0}^{\infty }dzV(\sqrt{2z})e^{-z}\left\{ \sum_{M=N,N+1}M%
\left[ {\rm L}_{M}(z){\rm L}_{N}^{1}(z)-{\rm L}_{M-1}(z){\rm L}_{N-1}^{1}(z)%
\right] \right\} =\frac{e^{2}\sqrt{2\pi }}{\chi \lambda }\theta _{N+1}, 
\]
was calculated for each odd-integer filling $\nu =2N+1$ ($N=0,1,2,...$); $%
{\rm L}_{N}^{m}(z)$ are the generalized Laguerre polynomials. For $%
V(r)=e^{2}/r\chi $, which is a reasonable approximation for the 2D electron
interaction in a narrow QW $\left( w\ll \lambda \right) $, $\theta _{1}=%
\frac{1}{4}$, $\theta _{2}=\frac{7}{16}$, $\theta _{3}\approx
0.\,\allowbreak 57$, $\theta _{4}\approx 0.\,\allowbreak 67$, and $\theta
_{5}\approx 0\allowbreak .\,\allowbreak 76$. Since the relevant part of the
SO coupling is off-diagonal with respect to the Landau level number, it
appears only in the second-order of a perturbation theory, or due to its
mixing with the $H_{{\rm \Omega }}$-term \cite{IordF}. The single-particle
spin-splitting, $\varepsilon _{{\rm Z}}$ in Eq. (\ref{Phi}) is corrected by
the effect of the 2D single-particle SO-coupling: $\varepsilon _{{\rm Z}%
}=\mu gB-\alpha \nu \left( \widetilde{u}_{{\rm so}}^{2}-u_{{\rm so}%
}^{2}\right) /\omega _{{\rm c}}\lambda ^{2}$. We also include into $\Phi (%
{\bf n},\nabla {\bf n})$ the topological term and the Coulomb energy of
additional charges, $\rho ({\bf x})$, in order to discuss charged
skyrmion-type textures, 
\begin{equation}
E_{{\rm sk}}=\frac{\Im _{N+1}}{2}\rho +\int d{\bf x}^{\prime }\rho ({\bf x})%
\frac{V\left( {\bf x}-{\bf x}^{\prime }\right) }{2}\rho ({\bf x}^{\prime
});\;\rho ({\bf x})=\frac{-\alpha }{8\pi }\epsilon ^{\beta \gamma \delta
}\epsilon ^{ij}n^{\beta }\partial _{i}n^{\gamma }\partial _{j}n^{\delta }.
\label{density}
\end{equation}

\ifpreprintsty 
\else
{\tiny\hspace*{\fill}\begin{tabular}[t]{|c}\hline
\parbox{0.49\hsize}{~} \\
\end{tabular}}\vspace*{-2.5ex}%
\begin{multicols}{2}\noindent
\fi

From a phenomenological point of view, thermodynamic potential $\Phi \left\{ 
{\bf n}({\bf x})\right\} $ in Eq. (\ref{Phi}) describes an easy-axis
ferromagnet with square 2D Bravais lattice and broken inversion symmetry,
and in a perpendicular magnetic field. It contains all terms in the
magnetization energy allowed by the crystalline symmetry of the $\left[ 001%
\right] $-grown quantum well in a zinc-blend-type semiconductor \cite{RemDiv}%
. The sketched above microscopic derivation of $\Phi \left\{ {\bf n}({\bf x}%
)\right\} $ was necessary to obtain the values of coefficients in front of
the phenomenologically allowed invariants composed of ${\bf n}$ and $\nabla $%
. The first term in $\Phi $ describes spin-stiffness. The second term
determines an easy-axis anisotropy along ${\bf l}_{+}=\left[ 110\right] /%
\sqrt{2}$. For structures with $\widetilde{u}_{{\rm so}}\gg u_{{\rm so}}$,
it rather defines a weakly anisotropic easy plane for spin polarization,
which competes with the Zeeman energy term. Such a competition resumes
itself in the deviation of polarization from a fully ${\bf l}_{z}$-aligned
state at $\left| \varepsilon _{{\rm Z}}\right| <\varepsilon _{{\rm Z}}^{{\rm %
e}}$, 
\begin{equation}
\varepsilon _{{\rm Z}}^{{\rm e}}=2\left( u_{+}/\lambda \omega _{{\rm c}%
}\right) ^{2}\Im _{N+1},\;u_{\pm }=\widetilde{u}_{{\rm so}}\pm u_{{\rm so}}.
\label{easyc}
\end{equation}
As a function of a varying $\varepsilon _{{\rm Z}}$, this can be viewed as a
second order phase transition into the easy-axis state, 
\begin{equation}
{\bf n}_{\pm }=-\left( \varepsilon _{{\rm Z}}/\varepsilon _{{\rm Z}}^{{\rm e}%
}\right) {\bf l}_{z}\pm {\bf l}_{+}\sqrt{1-\left( \varepsilon _{{\rm Z}%
}/\varepsilon _{{\rm Z}}^{{\rm e}}\right) ^{2}}  \label{easyn}
\end{equation}
across which the symmetry between $\pm {\bf l}_{+}$ magnetic directions gets
spontaneously broken \cite{Advice}.

Since for $\left| \varepsilon _{{\rm Z}}\right| <\varepsilon _{{\rm Z}}^{%
{\rm e}}$ both easy-axis configurations, ${\bf n}_{+}$ and ${\bf n}_{-}$
have the same energy density, 
\begin{equation}
\Phi _{{\rm e}}=-\frac{\Im _{N+1}}{4\pi \lambda ^{2}}\left( \frac{%
u_{+}/\lambda }{\omega _{{\rm c}}}\right) ^{2}\left[ 1+\left( \frac{%
\varepsilon _{{\rm Z}}}{\varepsilon _{{\rm Z}}^{{\rm e}}}\right) ^{2}\right]
,  \label{energyEasy}
\end{equation}
the easy-axis state of a QHF tends to acquire a domain structure: by
splitting dynamically into the set of mesoscopic-size regions with opposite
polarization projections onto the $\left[ 110\right] $-axis. The latter
possibility has to affect the skyrmion-dominated dissipative transport
properties of a liquid. The matter is that the activation energy of a
skyrmion-antiskyrmion pair confined to the domain wall is lower, than in the
2D bulk \cite{IordF}. In fact, the larger is the difference in the
polarization between two domains, the more skyrmion is energetically
confined to it. Since ${\bf n}_{\pm }$-polarized states are degenerate, the
areas covered by ${\bf n}_{+}$ and ${\bf n}_{-}$ domains are statistically
equal, so that the network of better conducting domain walls (with a lower
activation energy for thermally excited carriers) forms a percolation
cluster, thus resulting in a continuous decline in the activation energy for
the macroscopic $\sigma _{xx}$\ which would follow the decrease of $%
\varepsilon _{{\rm Z}}$.

A further analysis of the functional in Eq. (\ref{Phi}) extended onto the
limit of $\varepsilon _{{\rm Z}}=0$ shows that, apart from the domain
structure formation, there is another reason for the field ${\bf n}({\bf r})$
to be inhomogeneous. The fourth term \cite{RemDiv} in Eq. (\ref{Phi}) tends
to twist the polarization field of a QHF into helical texture, 
\begin{equation}
{\bf n}({\bf r})={\bf l}\sin (\phi {\bf (r)})+{\bf l}_{z}\cos (\phi {\bf (r)}%
).  \label{helical}
\end{equation}
The latter is characterized by the helicity plane built upon two unit
vectors, ${\bf l}=(l^{x},l^{y},0)$ and ${\bf l}_{z}$, spatial orientation $%
{\bf m}$, and period ${\cal L}$, $\phi {\bf (r+m{\cal L})=}\phi {\bf (r)}%
+2\pi $. As a variational approximation, one can use ${\bf n}({\bf r})$ in
Eq. (\ref{helical}) with $\phi ({\bf r})={\bf mr/{\cal L}}$, treating ${\bf %
l,m}$ and ${\bf {\cal L}}$ as minimization parameters. The resulting texture
can be viewed as an image of a spoke in a wheel rolling in the direction of $%
{\bf m}$, with $\phi $ being the integral angle encircled by a spoke, so
that we shall call $\phi $ the 'helicity phase'. The energy density of an
optimal variational state, 
\begin{eqnarray}
{\bf m} &=&{\bf l=l}_{-}{\bf \equiv }[1\bar{1}0]/\sqrt{2},\;{\cal L}^{{\rm v}%
}=\pi \lambda ^{2}\omega _{{\rm c}}/u_{+},  \label{variational} \\
\Phi _{{\rm h}}^{{\rm v}} &=&-\frac{\Im _{N+1}}{4\pi \lambda ^{2}}\left\{
\left( \frac{u_{+}/\lambda }{\omega _{{\rm c}}}\right) ^{2}+\frac{1}{2}%
\left( \frac{u_{-}/\lambda }{\omega _{{\rm c}}}\right) ^{2}\right\} ,
\label{helicalen}
\end{eqnarray}
is lower than the energy $\Phi _{{\rm e}}(\varepsilon _{{\rm Z}}=0)$ of a
homogenous easy-axis configuration in Eq. (\ref{energyEasy}). Optimal
variational state provides us with values of $\Phi _{{\rm h}}$ and ${\cal L}$
which are very close to the ${\bf n}$-field distribution that really
minimizes \cite{RemFullSolution} the functional $\Phi $ in Eq. (\ref{Phi}).
For the experimentally relevant example of $u_{+}\approx u_{-}$ ({\it i.e.}, 
$\widetilde{u}_{{\rm so}}\gg u_{{\rm so}}$), $\Phi _{{\rm h}}^{{\rm v}}=-%
\frac{3}{2}\left( \widetilde{u}_{{\rm so}}/\lambda \omega _{{\rm c}}\right)
^{2}\left( \Im _{N+1}/4\pi \lambda ^{2}\right) $ and ${\cal L}^{{\rm v}}$ in
Eq. (\ref{variational}) should be compared \cite{RemFullSolution} with $\Phi
_{{\rm h}}\approx -1.53\left( \widetilde{u}_{{\rm so}}/\lambda \omega _{{\rm %
c}}\right) ^{2}\left( \Im _{N+1}/4\pi \lambda ^{2}\right) $ and ${\cal L}%
\approx 3.25\omega _{{\rm c}}\lambda ^{2}/\widetilde{u}_{{\rm so}}$. We,
therefore, assess the stability of a helical state, against the easy-axis
one, on the basis of the energetics of variational helical texture with
parameters given by Eq. (\ref{variational}).

Due to the difference in the symmetry of a helical and easy-axis states,
which cannot be continuously transformed one into another, the
transformation between them can be viewed as a first-order phase transition.
The condition for such a transition, $\Phi _{{\rm h}}=\Phi _{{\rm e}%
}(\varepsilon _{{\rm Z}})$, determines critical value of Zeeman splitting, $%
\varepsilon _{{\rm Z}}^{{\rm h}}$ estimated as 
\begin{equation}
\varepsilon _{{\rm Z}}^{{\rm h}}=\sqrt{2}\left( u_{+}u_{-}/\omega _{{\rm c}%
}^{2}\lambda _{B}^{2}\right) \Im _{N+1}\approx \varepsilon _{{\rm Z}}^{{\rm e%
}}/\sqrt{2}.  \label{helicalc}
\end{equation}

The helical state formation can manifest itself in several observations. For
example, the local value of the Knight shift \cite{Barrett}, $\delta _{{\rm %
hf}}$ in the spin-splitting of Ga and As nuclei located in the QW acquires
an alternate coordinate-dependent sign, thus modifying the NMR lineshape, $%
I(\delta )$. Locally, the NMR shift $\delta =\left( \omega -\omega
_{0}\right) /\delta _{{\rm hf}}$ is due to the hyperfine interaction of
nuclear spins with fully polarized electrons, with $\delta _{{\rm hf}}$
being its maximal value just in the QW center. In a homogeneously polarized
gas, the NMR line from the QW has a double-peak structure \cite{Barrett}, $%
I(1>\delta >0)\propto \left[ \delta (1-\delta )\right] ^{-1/2}$, with a
distinct satellite at $\delta =1$ split by the hyperfine coupling. In the
easy-axis configuration, Eq. (\ref{easyn}), the double-peak structure
persists, with a reduced maximal splitting: $\delta _{{\rm hf}}\rightarrow
\left( \varepsilon _{{\rm Z}}/\varepsilon _{{\rm Z}}^{{\rm e}}\right) \delta
_{{\rm hf}}$, as far as $n^{z}=-\varepsilon _{{\rm Z}}/\varepsilon _{{\rm Z}%
}^{{\rm e}}$. On the contrary, in a helical phase, this has to transform
into a single broadly tailed resonance with a non-Lorenzian shape
approximated by $I(\delta )\propto \left| \delta \right| ^{-1/2}$ for $%
1>\delta >-1$.

The anisotropy of transport characteristics of a QHF with respect to $[110]$
and $[1\bar{1}0]$ crystallographic directions may be another feature of the
helical state. Speaking about dissipative conductivity formed by thermally
activated electron-hole pairs at the spin-split LL's, this can be understood
after taking into account that charge-carrying excitations determined in a
locally rotated spin-frame (adjusted to ${\bf n}({\bf r})$) are subjected to
a smooth (${\cal L}\gg \lambda $) potential due to Zeeman energy, $\left(
\varepsilon _{{\rm Z}}/2\right) \cos ([x-y]/\sqrt{2}{\cal L})$. Therefore,
at low temperatures, $T<\varepsilon _{{\rm Z}}$ the dissipative conductivity 
$\sigma _{-,-}$ along ${\bf m\parallel }[1\bar{1}0]$ would be suppressed, as
compared to $\sigma _{+,+}$ (across ${\bf m}$).

For the dissipative transport dominated by skyrmions \cite{Eisenstein}, the
difference between $\sigma _{-,-}$ and $\sigma _{+,+}$ is to be the result
of the anisotropy of a skyrmion itself. In fact, the form of a skyrmion in a
helically twisted texture is quite complex: In a periodic system, these are
dislocations which represent the very elementary topological defects, rather
than skyrmions. The periodicity of a helical texture in Eq. (\ref{helical})
is controlled by the helicity phase $\phi _{0}({\bf mr/{\cal L}})$ in Eqs. (%
\ref{helical}) and (\ref{variational}). One missing (or extra) period in one
half of a plane, as compared to the other half (a dislocation) is equivalent
to the phase shift of $\pm 2\pi $\ accumulated at large distances from the
dislocation core, thus resulting in the winding number $D=\pm 1$. On the
other hand, we assume that the dislocation core is not singular. To
illustrate the topology of a non-singular core, let us draw a large-radius
circle around a dislocation. At large distances, where helical structure is
not perturbed, such a contour maps into the equator of a unit sphere and
encircles it $N$ or $\left( N-1\right) $ times, depending on which one of
two semi-circles is retraced: drawn above, or below the dislocation.

Upon moving the upper half of a contour down through the dislocation, an
extra loop encircling the unit sphere equator should continuously disappear.
The latter is possible if the contour image slips through either the $+{\bf l%
}_{+}$, or, alternatively, $-{\bf l}_{+}$ pole of a unit sphere, which can
be modelled by such a field configuration ${\bf n(r)}={\bf l}_{z}n^{z}+{\bf l%
}_{+}n^{+}+{\bf l}_{-}n^{-}$, that 
\[
n^{z}+in^{-}=e^{i\phi +iD\varphi }\sqrt{1-\left( n^{+}\right) ^{2}},\left\{ 
\begin{array}{c}
n^{+}(0)=\pm 1, \\ 
n^{+}(r\rightarrow \infty )=0,
\end{array}
\right. 
\]
where $(r,\varphi )$ are polar coordinates calculated from the dislocation
center, $r=0$. The image of a 2D plane provided by ${\bf n(r)}$\ maps onto
only one half of a sphere $\left| {\bf n}\right| =1$, so that the
dislocation core is characterized by an additional topological number, $%
\vartheta \equiv n^{+}(0)=\pm 1$ distinguishing between 'left' and 'right'
semi-spheres. Using $\rho ({\bf x})$ in Eq. (\ref{density}), we find that
the core of a dislocation ($D=1$) or anti-dislocation ($D=-1$) carry a
half-integer electric charge $\int d{\bf x}\rho ({\bf x})=\frac{1}{2}%
\vartheta D$. However, an isolated dislocation has a logarithmically large
energy, 
\[
{\cal E}\left( D=\pm 1\right) =\int \frac{d{\bf x}}{2\pi }\frac{\Im _{N+1}}{8%
}\left( \nabla \varphi \right) ^{2}\approx \frac{\Im _{N+1}}{8}\ln (r/{\cal L}),
\]
and, at low temperatures, dislocations and anti-dislocations have to form
pairs bound by a long-range (logarithmic) attraction, except, maybe, for a
possible Kosterlitz-Thouless melting effect. Since both dislocation and
anti-dislocation in a bound pair may be equally charged,\ such \ a pair, ($%
D,\vartheta $) and ($-D,-\vartheta $) together constitute a skyrmion.

The result of the above analysis of phases of a ferromagnetic quantum Hall
effect liquid in a narrow QW can be summarized as follows. Upon decreasing
the single-particle Zeeman splitting, $\left| \varepsilon _{{\rm Z}}\right| $
({\it e.g.}, by pressure), spin polarization of a liquid starts to acquire
at $\left| \varepsilon _{{\rm Z}}\right| =\varepsilon _{{\rm Z}}^{{\rm e}}$
the easy-axis configuration, which is followed by the abrupt fall into a
helical state at $\left| \varepsilon _{{\rm Z}}\right| =\varepsilon _{{\rm Z}%
}^{{\rm h}}\approx \varepsilon _{{\rm Z}}^{{\rm e}}/\sqrt{2}$. Using the
bulk SO coupling parameter $\gamma =25$eV\AA $^{3}$, as a reference, we
estimate for the $\nu =1$ liquid in a $68$\AA -wide GaAs/AlGaAs quantum well
structure with a carrier density $2.8\times 10^{11}$cm$^{-2}$ studied by
Maude {\it et al} \cite{Maude} that $\varepsilon _{{\rm Z}}^{{\rm e}%
}/(e^{2}/\chi \lambda )\approx 1.5\times 10^{-3}$, and $\varepsilon _{{\rm Z}%
}^{{\rm h}}/(e^{2}/\chi \lambda )\sim 1\times 10^{-3}$, which roughly fits
into the range of a variable Zeeman energy in Ref. \cite{Maude}, where the
dissipative transport activation has been drastically affected by pressure.
The helical texture period estimated for the same parameters is ${\cal L}%
\sim 5\times 10^{3}$\AA . Note that, according to Eqs. (\ref{helicalc}) and (%
\ref{easyc}), the parametric range of pressures where helical and easy-axis
phases are stable is broader for higher odd-integer filling factors in the
same density structure.

Authors thank N.Cooper, P.Littlewood, A.MacDonald, Yu.Nazarov, R.Nicholas,
M.Potemski and G.Volovik for discussions. This work has been supported by
EPSRC, INTAS and NATO.

\end{multicols} 


\begin{references}
\bibitem{Advice}  Our attention to the relevance of the easy-axis anisotropy
has been turned by Yu.Nazarov and by N.Cooper.
\bibitem{RemDiv}  This expression for the energy density is given neglecting 
$\left( \Im _{N+1}/2\omega _{{\rm c}}\lambda _{B}^{2}\right) {\rm div}\left[
u_{{\rm so}}\left( \alpha -n^{z}\right) {\bf n}_{\Vert }+\widetilde{u}_{{\rm %
so}}\left( \alpha +n^{z}\right) \widetilde{{\bf n}}_{\Vert }\right] $, which
can be transformed into the edges term.
\end{references}
\end{document}